\documentclass[11pt,dvips]{article}

\input{epsf}
\usepackage{wrapfig}
\usepackage{floatflt}
\usepackage{psfig}
\makeatletter
\renewcommand{\section}{\@startsection{section}{1}{0in}
	{0.4\baselineskip}{0.1\baselineskip}{\Large\bf}}
\renewcommand{\subsection}{\@startsection{subsection}{2}{0in}
	{0.25\baselineskip}{-\baselineskip}{\large\bf}}
\renewcommand{\subsubsection}{\@startsection{subsubsection}{3}{0in}
	{0.1\baselineskip}{-\baselineskip}{\normalsize\bf}}

\makeatother
{\catcode`\@=11
\gdef\SchlangeUnter#1#2{\lower2pt\vbox{\baselineskip 0pt\lineskip0pt
\ialign{$\m@th#1\hfil##\hfil$\crcr#2\crcr\sim\crcr}}}}

\begin{document}

%
\thispagestyle{myheadings}
%
%

\markright{OG 2.2.10} 

\begin{center}
{\LARGE {\bf Diffusive Shock Acceleration of Electrons and\\[0pt] Radio Emission from Large Diameter Shell-Type\\[0pt] Supernova Remnants\\}}
\end{center}


\begin{center}
{\bf A.I.Asvarov }\\[0pt]
{\it Institute of Physics, Azerbaijan Academy of Sciences, Baku, 370143, AZERBAIJAN
}\\[4pt]


{\large {\bf Abstract\\[0pt]
}}
\end{center}

\vspace{-0.5ex}
In present study I examine the capability of diffusive shock acceleration
mechanism to explain existing data on radio emission from evolved large
diameter shell-type adiabatic supernova remnants (SNRs). Time-dependent
''onion-shell'' model for the radio emission of SNRs is developed, which is
based on the assumptions: a) acceleration takes place from thermal energies
and test-particle approximation is valid; b) the problem of injection is
avoided by introducing, like Bell (1978), two injection parameters; c) to
take into consideration very late stages of SNR evolution the analytic
approximation of Cox and Andersen (1982) for the shell structure is used; c)
no radiative cooling. Constructed Surface Brightness - Diameter $(\Sigma -D)$
tracks are compared with the empirical $\Sigma -D$ diagram. The main
conclusion of the study is that the DSA mechanism is capable of explaining
all the statistics of radio SNRs including very large diameter remnants and
giant galactic loops 
\vspace{1ex} 

\section{Introduction:}

\label{intro.sec} Diffusive shock acceleration (DSA) mechanism are believed
to be the source of radio emitting relativistic electrons in young and
middle age SNRs of diameter $D\leq 20$ {\rm pc}. But for the radio emission
of evolved large diameter SNRs the mechanism of van der Laan (1962) or some
modification of this mechanism (e.g., Blandford and Cowie 1982) are widely
believed to be responsible. According to these models preexisting in the
ambient ISM electrons compressed to high level at the radiative shock are
responsible for the radio emission of the remnant. Due to the unstability of
radiative shock waves this mechanism may however encounter a number of
difficulties in reproducing the radio emission from very large diameter SNRs
evolving in warm phase of the ISM. Moreover, if the density of the ambient
medium is very low, the SNR will finish its life by merging with the ISM
before cooling becomes important. For such remnants DSA becomes a main
candidate for generation of relativistic electrons radio emitting in the
magnetic fields of $10^{-5}\div 10^{-6}$ {\rm G}. But electron acceleration
in SNRs is more difficult to estimate quantitatively since the injection
efficiency and even the very process of acceleration for electrons are still
unclear.

The goal of present study is to apply DSA mechanism, under very simple and
common assumptions about the injection, to follow the evolution of
shell-type SNRs up to the very large radii and, at the same time, to obtain
some new constraints on the theory of acceleration mechanism if we will be
able to achieve agreement between the model predictions and observations.

\section{The Model:}

\label{format.sec} We used the onion-shell model of Moraal and Axford (1983)
as it was done in Asvarov (1992, 1994) where the Sedov solution for the
remnant structure had been adopted. In present study SNR is modeled by using
an analytical approximation of Cox and Anderson (1982) which follows the
development of an adiabatic spherical blast wave in homogeneous ambient
medium of finite pressure. At early times this approximation resembles the
zero pressure Sedov similarity solution but extends the range of
investigation well into the regime in which the external pressure is
significant.

To avoid the problem of injection we introduced, like Bell (1978), two
injection parameters: we inject electrons as ``test particles'' at momentum $%
p_{{\rm inj}}=\psi \,p_{{\rm th}}$ , where $\,p_{{\rm th}}=(2{\rm m}_{{\rm e}%
}T_{{\rm s}})^{1/2}$ is the downstream electron thermal momentum, the
concentration of which assumed to be proportional to the density of the
ambient thermal electrons: $N_{{\rm inj}}=\varphi \,n_{{\rm oe}}$. At the
shock front equipartition between electron and proton temperatures is
assumed which means that $p_{{\rm inj}}$ is proportional to the velocity of shock wave.

For the accelerated electrons from the energy loss processes only adiabatic
cooling was taken into consideration.

Assuming the magnetic field to be frozen in to the plasma, we model the
density dependence of $H$ as $H=H_{{\rm 0}}(\rho /\rho _{{\rm 0}})^{{\rm k}}$%
, where $H_{{\rm 0}}$ is the ambient value of the magnetic field strength.

Although in our analysis we consider only extended SNRs we calculate several
models in high density environments for which the radiative phase begins
relatively soon after the explosion of SN. In this case total flux is
obtained by integration over the layers of the shell where radiative cooling
does not occur. This implies that we completely ignore the action of DSA at
radiative shocks.

\section{Results and Discussion:}

\label{session.sec} Empirical $\Sigma -D$ relations are very useful tools
for testing the theoretical models of SNR evolution. Before comparing our
model with the observations, we formulate the common properties of the model
predictions. Here we concentrate on two main observable radio
characteristics of SNRs: the spectral index and surface brightness. DSA at
strong shock waves in test particle approximation predicts for spectral
index the value of $0.5$, which will increase as the shock intensity (Mach
number) decreases. Calculations show that at Mach numbers $M\leq 4$ the
value of the mean radio spectral index gets greater then $0.6$ but it
remains bounded by maximal value of $0.75$ during the following evolution.
This is the result of distribution of magnetic field strength at the shock
but the real average spectrum of electrons inside the remnant will be
somewhat softer than $2\alpha +1=2.5$.

What concerns another important radio characteristic of the SNR, the surface
brightness, the remnant evolves at nearly constant radio surface brightness
followed by very steep drop. It is important to note that the dependence of
these and other radio characteristics of the remnant on shock Mach number
has almost universal nature, very weakly depending on the input parameters.

To compare the model predictions with the observations we have collected a
set of shell-type and a few composite SNRs in our Galaxy with known
distances and remnants in LMC. Corresponding $\Sigma -D\,\,\,$diagram is
shown in Fig.1. Error bars are due to uncertainties in the distances for
some SNRs; for several objects only lower and for one SNR upper limits are
known.
\begin{figure}[t]
\centerline{\hskip 0.0truecm
             \psfig{figure=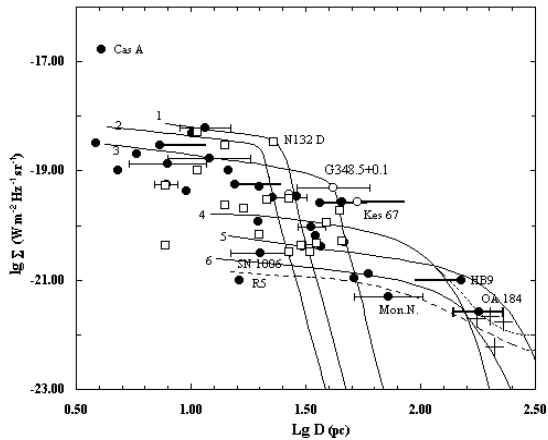,width=11cm}}

%
\caption{
Surface brightness - diameter diagram at 1 GHz for SNRs in Galaxy (circles)
and LMC (squares) and giant galactic loops (crosses). All the data for
galactic SNRs are taken from the catalogue of Green (1998), the data for
Loops are taken from Berkhuijsen (1986), and the data for SNRs in LMC are
from (Mills et al. 1984). Galactic SNRs with $\alpha \leq 0.4$ or/and
composite SNRs indicated by open circles. Curves are the modeled
evolutionary tracks: 1 - $n_{\rm{oe}}$=5, $H_{0}$=1, $E_{\rm{SN}}$%
=2; 2 - the same as 1 but $E_{SN}=1$; 3 -  $n_{\rm{oe}}=1$, $H_{0}=1$, $%
E_{\rm{SN}}=1$; 4 -  $n_{\rm{oe}}=0.05$, $H_{0}=0.5$, $E_{\rm{SN}%
}=2$; 5 -  $n_{\rm{oe}}=0.01$, $H_{0}=0.3$, $E_{\rm{SN}}=5$; 6 -  $%
n_{\rm{oe}}=5\,\,10^{-3}$, $H_{0}$ $=0.3$, $E_{\rm{SN}}=2$; 7 -  $n_{%
\rm{oe}}=5\,\,10^{-3}$, $H_{0}=0.3$, $E_{\rm{SN}}=1$. Here $E_{%
\rm{SN}}$ is in $10^{51}$ erg, $H_{0}$ is in $10^{-5}$ G, $n_{\rm{oe}}
$ is in cm$^{-3}$. Dashed curves correspond to models in which the
contribution of background cosmic ray electrons was considered.
}
\end{figure}

As a standard set of input parameters we used: the energy of SN explosion
$E_{\rm{SN}}$ which is varied in the range $(1\div 5)\,10^{51}$ ${\rm erg}$; the
ambient thermal electrons density $n_{{\rm 0e}}$ in the range $(5\div
5\,\,10^{-3})$ ${\rm cm}^{-3}$; the strength of the ambient magnetic field $%
H_{0}$ in the range$(3\div 10)\times 10^{-6}{\rm G}$. In all models the mass
of SN ejecta is taken to be one solar mass. In Fig.1 several $\Sigma _{1{\rm %
GHz}}(D)$ - tracks are shown for different values of $n_{{\rm oe}}$, $E_{%
{\rm SN}}$ and $H_{{\rm 0}}$. In all calculations we used $\psi =3$ and $%
\varphi =4\times 10^{-4}$ justifying the use of test particle approximation.
It is important to note that the shapes of the evolutionary tracks depend
very weakly on these parameters although the magnitude of $\,\Sigma $
depends on $\varphi$ linearly.

As can be seen in Fig.1 varying mainly the values of $n_{{\rm oe}}$ and $E_{{\rm SN}}$
the model is able to cover all the remnants in the $\Sigma -D$ diagram
including two large diameter relatively bright SNRs (HB 9, OA 184) and giant
radio loops. As the latter is attained with the price of somewhat large
value for $E_{{\rm SN}}$ of $5\times 10^{51}${\rm erg} we calculated several
models in which the contribution of the ambient relativistic electrons was
included via the injection of them into the DSA. As an example we have taken
the spectrum $j=1.4\times 10^{2}E^{-2.2}$ {\rm electrons m}$^{{\rm -2}}${\rm %
\ s}$^{{\rm -1}}${\rm \ sr}$^{{\rm -1}}${\rm \ GeV}$^{{\rm -1}}$ ($E$ in
{\rm GeV}) from Fichtel et al. (1991). In Fig.1 two tracks are drawn by
dashed lines.

It is interesting to not that the shapes of predicted by our model
evolutionary tracks are in excellent accordance with the prediction made by
Berkhuijsen (1986) that ''radio remnants may evolve adiabatically at nearly
{\it constant} $\Sigma _{{\rm R}}$, followed by a steep decrease''

As can be seen in Fig.1 radio evolution of SNRs depends on two parameters, $%
E_{{\rm SN}}$ and $n_{{\rm 0e}}$, equally.

In the framework of present model a number of features of the empirical $%
\Sigma -D\,\,\,$ relation obtains a simple explanation. For instance, the
small number of remnants with small diameters and low $\Sigma $ (lower left
corner in the diagram) is the result of very fast evolution of SN blast wave
in the low density ISM where SNRs have low $\Sigma $. According to our model
it is easy to account for high concentration of SNRs at diameters $30-50$ $%
{\rm pc}$ in the $\Sigma -D$ diagram: different kind of evolutionary tracks
intersect at these diameters and the sample of remnants here consists of
objects evolving at different initial conditions. Of course, in the origin
of the empirical $\Sigma -D$ relations we can not exclude at all the
contribution of various selection effects. Moreover, not all the remnants
can be described by our model. Indeed, in the $\Sigma -D$ diagram (Fig.1)
the composite SNRs and SNRs with $\alpha \leq 0.4$ (indicated by open
circles) have systematically large values of $\Sigma $ , which can be
understood that in these remnants an additional more effective mechanism
acts.

It is important to note that adopted values for Bells' parameters do
not contradict the observations at standard values for the input parameters,
characterizing the ISM and the SNR itself. This fact implies that the test
particle approximation has factual realization in evolved shell-type SNRs.
One more argument which favors the DSA mechanism is the statistics of
spectral indices. The catalogue of SNRs of Green (1998) contains 80 SNRs
with well determined values of the spectral indices,$\alpha $, from which $57
$ remnants $(71\%)$ have $\alpha \geq 0.45$ and only two SNRs (one of them
is young peculiar SNRs Cas A) $\alpha \geq 0.75$. Practically there are no
objects contradicting the prediction of our model that $\alpha _{{\rm \max }%
}\leq 0.75$. It is well known that young SNRs have systematically large
values of $\alpha $, which can be explained by the back reaction effects or by
the action of other then DSA mechanisms. According to the model SNRs
with Mach numbers $M\leq 4$ has the mean value of $\alpha \geq 0.6$. Assuming
for simplicity that SNR evolves according to the Sedov law, $M\propto
t^{-3/5}$, for the number of SNRs with Mach numbers greater then $M$ we have
$N(\geq M)\propto M^{-5/3}$ from which it follows that the number of SNRs
with $\alpha \geq 0.6$, $N(\alpha \geq 0.6)=N(M\leq 4)$, must be about $55\%$
of total number of SNRs. Here we have adopted for the final Mach number $M_{%
{\rm f}}=2.5$ at which $\Sigma $ drops more then two order of its initial
value and the SNR becomes invisible. In the catalogue Green (1998) $22$ out
of $80$ SNRs have $0.60\leq \alpha \leq 0.75$ which makes $27.5\%.$ This
discrepancy easily can be explained by the selection effects that SNRs with
large diameters and small Mach numbers have low $\Sigma $, consequently,
difficult to be detected, though their number is more than the number of
bright remnants by a factor of $8-10$.

The mean size and age of SNR with $E_{{\rm SN}}=10^{51}{\rm ergs}$, evolving
in the ISM with $n_{{\rm 0e}}=0.01\,\,{\rm cm}^{-3}$, when $\Sigma $ drops
to $\sim ~5\times 10^{-22}\,\,({\rm Wm}^{{\rm -2}}{\rm sr}^{{\rm -1}}{\rm Hz}%
^{{\rm -1}})$ are $150$ {\rm pc} and $2\times 10^{5}$ {\rm years},
respectively. If SN occur with a rate of $30$ {\rm year}$^{{\rm -1}}$ then
such SNRs occupy $3.5\times 10^{65}{\rm cm}^{{\rm 3}}$ of the galactic
volume of $\pi \times (25{\rm kpc})^{2}\times (1{\rm kpc})/6=9.65\times
10^{66}{\rm cm}^{3}$, or $1$ part in $28(3.6\%).$ This estimate depends on
the value of $E_{{\rm SN}}$ as $E_{{\rm SN}}^{4/3}$. The probability that
random line of sight will hit such SNR is $0.21$ or $10$ in $47$ and the
dependence on $E_{{\rm SN}}$ is the same. The last estimate shows that SNRs can
play important role in the origin of background radio and gamma emissions of
our Galaxy. Predicted by our model spectral indices are in accordance with
the radio background observations. What concerns the gamma-background our
model in accordance with the model of ''bubbling swiss cheese'' of Pohl \&
Esposito (1998) thought the value of electron spectral index of $2.0$
demanded in their model is in contradiction with our predictions.

\section{Conclusions:}

The model based on the assumption that the radio emitting electrons are
accelerated by DSAmechanism very well explains the statistics of shell-type
radio SNRs. From this we can conclude that the test particle approximation
and the supposition that acceleration of electrons takes place from the
thermal energies has realization in evolved SNRs in spite of all
theoreticaldifficulties concerning the physics of collisionless shock waves.

We obtain that $\psi =p_{{\rm inj}}/p_{{\rm th}}\simeq 3$. The idea that
there is no acceleration at the radiative stage of SNR evolution does not
contradict observation.

Presented model also in accordance with radio and gamma background
observations.
\vspace{-3pt}
\vspace{1ex}
\begin{center}
{\Large\bf References}
\end{center}
\vspace{-3pt} 
Asvarov, A.I. 1992, AZh 69, 753\newline
Asvarov, A.I. 1994, AZh 71, 228\newline
Bell, A.R. 1978, MNRAS 182, 443\newline
Berkhuijsen, E.M. 1986, A\&A 166, 257\newline
Blandford, R.D. \& Cowie, L.L. 1982, ApJ 260, 625\newline
Bogdan, T.J.\& Volk, H.J. 1983, A\&A,122,129\newline
Cox, D.P.\& Anderson, P.R. 1982, ApJ 253, 268\newline
Fichtel, C.E., Ozel, M.E., Stone, R.G.,\& Sreekumar, P., 1991, ApJ 374,134%
\newline
Green, D.A., 1998, ' A Catalogue of Galactic Supernova Remnants (1998

September version)', MRAO, Cambridge, UK\newline
Mills, B.Y., Turtle, A.J., Little, A.C. \& Durdin, J.M., 1984, Austr.J.Phys.

37, 321\newline
Moraal, H.\& Axford, W.I, 1983, A\&A 125, 204\newline
Pohl, M.\& Esposito, J.A. 1998, Preprint LANL, Astro-ph 9806160\newline
van der Laan, H. 1962, MNRAS 124, 179\newline

\end{document}